# The Vavilov-Cherenkov radiation by relict photon gas

S.G. Chefranov

A.M. Obukhov Institute of Atmospheric Physics , Russian Academy of Sciences,

Moscow, Russia; schefranov@mail.ru

Based on the new quantum Vavilov-Cherenkov,s radiation(VCR) theory (PRL,2004),it is stated the possibility of threshold realization in the modern epoch of the novel VCR effect by the relict photon gas, when it is traversed by relativistic particles of the cosmic rays with $\gamma \geq \gamma_{th} \cong 2 \cdot 10^{10}$, where $\gamma^{-2} = 1 - \frac{v^2}{c^2}$, v is the speed of particles, and c is the speed of the light in vacuum. Estimate of the number of radiated Cherenkov photons is obtained which (contrary to the estimates basing on the known standard VCR theory) is already sufficient for the possibility of viewing.

After the discovery of a relict background cosmic radiation, various mechanisms of its interaction with relativistic particles of the cosmic rays were actively investigated [1]. In relation with the observations of the particles with energy $E > 10^{20} eV$, in [2, 3] there was discussed also the possibility of the Vavilov-Cherenkov radiation (VCR) effect realization, when describing such an interaction on the basis of the standard VCR theory [4-6]. In this case, however, in [2, 3] it was obtained that for the observed in the modern epoch density of energy of the relict photon gas $(\varepsilon_{rel} \cong 10^{-12}\, erg/sm^3)$ corresponding refraction index n of the photon gas is found to be very close to one $(n - 1 \cong 10^{-42})$ [3], thus making impossible VCR effect due to un-really great value of the particles threshold velocity $v_{th} = c/n$, according to [4-6]. So in [2] for VCR it is required that $\gamma \geq \gamma_{th} \cong 10^{24}$, and in [3] $\gamma \geq \gamma_{th} \cong 10^{21}$, that is not compatible with the known GZK cutting [7,8], forbidding values $\gamma > 10^{11}$, according to the data of observation of spectrum of energy of cosmic rays particles.

In the present work, on the basis of the new quantum VCR theory [9, 10] (where, in particular, it is obtained $v_{th} = c/n_*$ for $n_* = n + \sqrt{n^2 - 1}$, when $n > 1$), it is shown that VCR effect by the relict photon gas is still possible in the modern epoch for $\gamma \geq \gamma_{th} \cong 2 \cdot 10^{10}$. In contrast to the quantum VCR theory of V.L. Ginzburg [6], in [9, 10] the role of a medium as a direct emitter of the VCR (see [4; 11]), is accounted not only in the equation of balance of the momentum but in the energy balance equation as well. In the result, it is obtained more accurate quantitative agreeing the estimate from [9, 10] of $v_{th}$ (in comparison to $v_{th}$ in [4-6] – see table in [9, 10]) with observed in experiment threshold of realization of VCR effect.

1. On the basis of the equations of balance of energy and momentum in [9, 10], the following definition of VCR threshold is obtained:

$$1 \geq \cos\theta = \frac{c}{vn_*}\left[1 + \frac{\varepsilon}{\gamma}\begin{cases}\dfrac{\sqrt{n^2-1}}{n}, n > 1 \\ \sqrt{1-n^2}, n < 1\end{cases}\right], \tag{1}$$

$$n_* = \begin{cases} n + \sqrt{n^2-1}, n > 1 \\ \dfrac{1+\sqrt{1-n^2}}{n}, n < 1 \end{cases},$$

where n is the medium refraction index, $\varepsilon = \frac{\varepsilon_p}{mc^2}$, $\varepsilon_p$ is energy of Cherenkov photon emitted under the angle θ relative to the direction of the particles motion, m is the rest mass of the particle moving with the constant speed v.

For (1) getting, in [9, 10] it is used an idea that when the medium emits a photon with energy $\varepsilon_p$, the energy of the medium shall be decreased by $m_p c^2$, where $m_p$ is the rest mass of a photon in a medium with $n \neq 1$. It is defined from the relativistic relation $m_p^2 c^2 = \frac{\varepsilon_p^2}{c^2} - p^2$, if Abraham representation [12] for the photon impulse value p in medium is used, when

$$p_a = \frac{\varepsilon_p}{c}\begin{cases}\frac{1}{n}, n>1\\ n, n<1\end{cases} \qquad (2)$$

In this case from (2), it follows that for any $n \neq 1$ the value of $m_p$ is finite and real, having the following form:

$$m_p = \frac{\varepsilon_p}{c^2}\begin{cases}\frac{\sqrt{n^2-1}}{n}, n>1\\ \sqrt{1-n^2}, n<1\end{cases}, \qquad (3)$$

Let's note that the use in [6] of the Minkowski representation $p_m = \frac{\varepsilon_p n}{c}, n>1$ yields also finite $m_p$ (otherwise, see [9], it is impossible to get for $\cos\theta$ the formula (11) in [6]), which, however, is imaginary contrary to (3). It is known [12], that such used in [6] representation of the pseudo-impulse of the photon, contrary to the given above representation (2), is not the true photon impulse in the medium since it necessarily contains in its definition also the information on impulse transferred to the medium, when the photon appears in it. This reason and experimental evidence (also see [12]) of the truth, leading to (2), symmetrical Abraham representation for the electro-magnetic field energy-impulse tensor in the medium (that concluded nearly 70-year discussion between supporters of Abraham and Minkowski representations) gives the basis for using in the new theory of VCR [9, 10] the photon impulse in the medium just in the form of (2).

For finite value of $\varepsilon$ in (1) in the limit $n-1 << 1$ and $\gamma >> 1$, from (1) one gets the following definition of the threshold (for VCR realization) particle speed:

$$v > v_{th} = c/n_*, \qquad (4)$$

where value of $n_*$ is defined in (1). In the result, $v_{th}$ is found to be $n_*/n$ times less than the threshold speed in VCR theory [4-6].

2. Using the threshold VCR condition (4), one can get estimate $\gamma_{th}$, necessary for defining possibility of realization of VCR by relict photon gas in the moden epoch.

For this sake, it may be used , known from nonlinear quantum electrodynamics ,representation of n in the form [3,13,14]

$$n^2 = 1 + \delta, \quad \delta = \mathrm{kbQ}^2, \qquad (5)$$

where $\mathrm{k} = \dfrac{2\alpha^2\hbar^3}{45\mathrm{m}^4\mathrm{c}^5}$, $\alpha = \dfrac{\mathrm{e}^2}{\hbar\mathrm{c}}$ is the constant of thin structure, $\hbar$ is Plank constant, m and e are mass and charge of electron. In (5) $\mathrm{b} = 8$ or 14 – depending on the character of polarization of electromagnetic wave spreading in the photon gas. In particular, for the flat wave $\mathrm{Q}^2 = \frac{4}{3}\varepsilon_{\mathrm{rel}}$ [14] in (5), and in the modern epoch one gets the estimate [3] $\delta \cong 3 \cdot 10^{-42}$. For such small $\delta$ it is found out that an important for $v_{th}$ in (4) value $\mathrm{n}_* - 1 \cong \sqrt{\delta}$ on many orders exceeds the value of $\mathrm{n} - 1 \cong \frac{\delta}{2}$ (because. $\sqrt{\delta} >> \delta/2$ for $\delta << 1$), which is used in [2,3] to assess $v_{th}$ and the number of Cherenkov photons based on the standard theory [4-6].

For the given estimate (5) and based on (4), we get threshold value $\gamma_{\mathrm{th}} \cong 2 \cdot 10^{10}$, that yields the possibility of VCR realization by relict photon gas in the modern epoch for $\gamma > \gamma_{\mathrm{th}}$ without any contradiction now with the observed data and GZK cutting [7,8]. Obtained here threshold value by the order of value is found to be close to the specified GZK limit, noticed in [7, 8] without any relation to VCR. That’s why when interpreting corresponding observation data on the presence of the threshold cutting in the energy spectrum of the relativistic particles, it is necessary to take into account also explained in the new quantum theory [9, 10] threshold effect of VCR in line with GZK effect. Let’s note that the known standard theory of VCR [4-6] (see above, and [9] for more details) is not able to obtained the very threshold value of the VCR effect, when it gives good description of the already realized radiation, correctly defining only the place of the corresponding interferential maximum of VCR (but not the limit of the angle of the whole VCR cone). The new quantum VCR theory [9, 10] was used here not mainly for getting its (in addition to [9, 10]) experimental confirmation, but for the illustration of the simple possibility of its use in various regions of physics.

Important role plays, defining according to (1), anisotropy of VCR, which is absent for example in discussed in relation to cosmic rays (see, in particular, [15]) mechanism of the inverse Compton-effect. Thus obtained on the base of (1)-(5) conclusion on the reality of VCR in the modern epoch may be used for analyzing of observed data of cluster anisotropy of wide atmosphere showers [16,17] and anisotropy of cosmic background radiation [18].

3. Estimates of the number of radiated by the photon gas Cherenkov particles which were obtained in [2, 3] on the base of the VCR theory [4-6] were discussed above. Related to that, in [2], there is the following relationship (see formula (6) in [2]) between the number of Cherenkov photons $dN$ radiated in the frequency range $d\omega$ on the range $dl$ of the way of moving of a particle and the threshold (for the VCR possibility) particles energy $E_{th}$ (for $E_{th} >> mc^2$):

$$\frac{dN}{dl d\omega} = \frac{\alpha}{c(\frac{E_{th}}{mc^2})^2} \qquad (6)$$

For $n = 1 + \Delta n$ and $\Delta n << 1$, in [2], an estimate $E_{th} = mc^2 \sqrt{\frac{1}{2\Delta n}}$ is used which is obtained on the base of the VCR theory [4-6]. Note that in (6), the right-hand side is found to be proportional to the value of $\Delta n$ in the first power. For $\Delta n$ of the order of $10^{-48}$ for extremely high energy of VCR, in [2], it was obtained that on the way of the length 1Mpc ($3 \cdot 10^{24}$cm) the number of radiated Cherenkov photons is $N < 3 \cdot 10^{=5}$, i.e. it is not sufficient for the possibility of viewing them.

From the other side, if to use in (6) representation for $E_{th}$ obtained from the new quantum VCR theory [9, 10] (where according to (4) we have now $E_{th} = \frac{mc^2}{\sqrt{2\sqrt{2\Delta n}}}$ then under the same conditions we get the estimate, $N < 3 \cdot 10^{19}$ that is feasible for observing. Similarly, we can estimate N also in the case of low VCR energy corresponding to the representation (4) obtained in [3]. Now it is found out (for the case of the parameter values considered in [3]) that $N < 3 \cdot 10^{11}$.

Let's note that discussed above and in [9, 10, 19] new approach to VCR theory allows (see details in [9]) usage for interpretation of various astrophysical observations not only mechanism of synchrotron radiation (nowadays not having any rival) but VCR mechanism as well. It is specially important in the case of $n < 1$, when in distinction to the VCR theory [4-6], there is no ban (see (1)-(4)) on the possibility of radiation of transverse electromagnetic waves due to the direct realization of VCR mechanism.

## References

1. Ya.B. Zeldovich, I.D. Novikov, Structure and evolution of the Universe, Moscow, 1975.
2. I.M. Dremin, JETP Lett, **75**, 167 (2002)
3. M. Marklund, G. Brodin, L. Stenflo, P.K. Shukla, New Journal of Physics, **7**, 70 (2005)
4. U.E. Tamm, J. Phys. USSR, **1**, 439 (1939), Собр. науч. тр. т. 1, с.79, М. 1975
5. I.M. Frank, Uspehi Fizicheskih Nauk, **68**, 397 (1959)
6. V.L. Ginzburg, Uspehi Fizicheskih Nauk, **166**, 1033 (1996)
7. K. Greisen, Phys. Rev. Lett., **16**, 748 (1966)
8. G.T. Zatsepin, V.A. Kuzmin, JETP Lett., **4**, 78 (1968)
9. S.G. Chefranov, JETP, **126**, 333 (2004)
10. S.G. Chefranov, Phys, Rev, Lett., **93**, 254804 (2004)
11. L.D. Landau, Ye. M. Lifshits, Electrodynamics of dense matter, Moscow, 1982
12. V.L. Ginzburg, Theoretical physics and astrophysics, Moscow 1981
13. Z. Bialynicka-Birula, I. Bialynicki-Birula, Phys. Rev. **D2**, 2341 (1970)
14. M. Marklund, G. Brodin, L. Stenflo, Phys. Rev. Lett., **91**, 163601 (2003)
15. J.E. Felton, Phys. Rev. Lett., **15**, 1003 (1965)
16. A.V. Glushkov, JETP Letters, **75**, 3 (2002)
17. A.V. Glushkov, JETP, **136**, 893 (2009)
18. P. de Bernardis, et.al., Nature, **404**, 955 (2000)
19. S.G.Chefranov,arXiv: 1009.0594v1[astro-phHE] 6 Sep 2010